\documentclass[10pt,a4paper,english,english]{article}
\usepackage{ae,aecompl}
\usepackage[T1]{fontenc}
\usepackage[latin1]{inputenc}
\usepackage{amstext}
\usepackage{graphicx}
\usepackage{esint}

\makeatletter

\usepackage{babel}\usepackage{graphics}
\usepackage{multicol}

\usepackage{babel}

\usepackage{babel}

\usepackage{babel}

\makeatother

\usepackage{babel}
\begin{document}
\begin{flushright}
Proceedings of FEDSM 2006 \\
 {} 2006 ASME JOINT US European Fluids Engineering Summer Meeting\\
 {} July 17-20, 2006, Miami, FL, USA\\
{\large FEDSM2006-98383}{\LARGE{} }
\par\end{flushright}{\LARGE \par}

\vspace{10sp}

\begin{center}
{\large IN SEARCH OF RANDOM UNCORRELATED PARTICLE MOTION (RUM) IN
A SIMPLE RANDOM FLOW FIELD } 
\par\end{center}

\vspace{7.5sp}

\begin{multicols}{2}

\begin{center}
{\large Michael W. Reeks}\\
 {\Large {} }School of Mechanical \& Systems Engineering\\
 Stephenson Building, Claremont Rd\\
 {} University of Newcastle upon Tyne\\
 {} Newcastle upon Tyne, NE1 7RU, UK\\
 {\Large {} }{\large Luca Fabbro \& Alfredo Soldati}\\
 {\large {} }Department of Energy Technology \\
 {} University of Udine, Via delle Scienze 208 \\
 {} 33100 Udine,Italy\\
 {\large {} } 
\par\end{center}

\end{multicols}

\begin{center}
email: mike.reeks@ncl.ac.uk, soldati@uniud.it, luca.fabbro@uniud.it 
\par\end{center}

\begin{multicols}{2}

\textbf{ABSTRACT}\\
 DNS studies of dispersed particle motion in isotropic homogeneous
turbulence \cite{key-1} have revealed the existence of a component
of random uncorrelated motion (RUM) dependent on the particle inertia
$\tau_{p}$(normalised particle response time or Stoke number). This
paper reports the presence of RUM in a simple linear random smoothly
varying flow field of counter rotating vortices where the two-particle
velocity correlation was measured as a function of spatial separation.Values
of the correlation less than one for zero separation indicated the
presence of RUM. In terms of Stokes number, the motion of the particles
in one direction corresponds to either a heavily damped $\left(\tau_{p}<0.25\right)$
or lightly damped $\left(\tau_{p}>0.25\right)$ harmonic oscillator.
In the lightly damped case the particles overshoot the stagnation
lines of the flow and are projected from one vortex to another (the
so-called sling-shot effect). It is shown that RUM occurs only when
$\tau_{p}>0.25$, increasing monotonically with increasing Stokes
number. Calculations of the particle pair spatial distribution function
show that equilibrium of the particle concentration field is never
reached, the concentration at zero separation increasing monotonically
with time. This is consistent with the calculated negative values
of the average Liapounov exponent (finite compressibility) of the
particle velocity field.

\section{{\normalsize INTRODUCTION}}

Turbulent structures play a crucial role in many particle/fluid processes
in the environment and industry: in powder production, combustion,
and the formation and growth of PM10 particulates in the atmosphere.
An area of much investigation is the mechanism for warm-rain initiation
and in particular the way droplet interaction with the small scales
of turbulence in clouds influence the droplet size distribution. The
current consensus is that the turbulence demixes the particles leading
to a much enhanced collision rate (i.e. much greater than that based
on passive scalar motion\cite{key-2},\cite{key-3}). Our only particular
interest in this subject is motivated by the processes of turbulent
agglomeration and deagglomeration of a aerosols released in a steam
generator tube rupture in a PWR. Early experiments and simulations\cite{key-11}
have shown that this demixing reaches a maximum when the particle
response time is typical of the timescale of the turbulent structure
(i.e particle Stokes numbers $\sim$ 1), the suspended particles being
observed to segregate into regions of high straining rate in between
the regions of vorticity. In addition Maxey and his co-workers\cite{key-4},\cite{key-9}
showed that the gravitational settling of particles in homogeneous
turbulence was enhanced due \emph{preferential sweeping} in the direction
of gravity as particles interweave through turbulent structures in
the flow. Since then there have been numerous studies to understand
and quantify this segregation process. Of particular note have been
the seminal studies by Collins et al.\cite{key-10} and Wang et al.\cite{key-55}
to quantify the influence of segregation on relative dispersion (two
particle dispersion) and on particle agglomeration and the recent
work of Bec \cite{key-14} who expressed the clustering of particle
in terms of its fractal dimension and showed how this was related
to the Liapounov exponents of the particle phase space distribution.
Even more recently Vassilicos\cite{key-15} has shown that the clustering
is strongly related to the acceleration stagnation points in the flow
as they are swept by the large scale motions of the turbulence.

This paper is not directly concerned with the clustering but with
an intrinsic property of the motion of inertial particles in flow
fields that are spatially random but smoothly varying. Simonin et
al. \cite{key-1} have observed that the spatial velocity field resulting
from the motion of suspended particles in DNS isotropic homogeneous
turbulence consists of two components: a smoothly (continuous) velocity
field that accounts for all particle-particle and fluid-particle two
point spatial correlations (they referred to this components as the
mesoscopic Eulerian particle velocity field (MEPVF)); and a spatially
uncorrelated component which we will refer to here as RUM%
\footnote{Simonin et al. refer to this component as the quasi-Brownian velocity
distribution because of its similarity to the spatially uncorrelated
velocity distribution of Brownian particles%
} (the component of random uncorrelated motion) whose contribution
to the particle kinetic energy increases as the particle inertia increased.
Simonin et al \cite{key-1} attribute this feature to the ability
of the particles with inertia to retain their memory of their interaction
with \emph{very distant, and statistically independent eddies in the
flow field}. The flow field in their DNS of homogeneous isotropic
turbulence as with real turbulence is complicated with many scales
of spatial and temporal existence. In our study, the results of which
we present here, we show the presence of RUM in a much simpler flow
field of counter rotating vortices and in so doing are able to examine
the properties of RUM in relation to the particle interactions with
simple structures. The flow field we have used has been used before
in studies of particle clustering \cite{key-54} and in particular
with the measurement of the particle compressibility and intermittency.
Our eventual aim is to show how these properties are related to the
occurrence of RUM

\section{{\normalsize DESCRIPTION OF CARRIER FLOW FIELD \label{sec:Description of random flow field}}}

\begin{figure}
\begin{minipage}[c]{0.49\textwidth}%
\begin{center}
\includegraphics[scale=0.45]{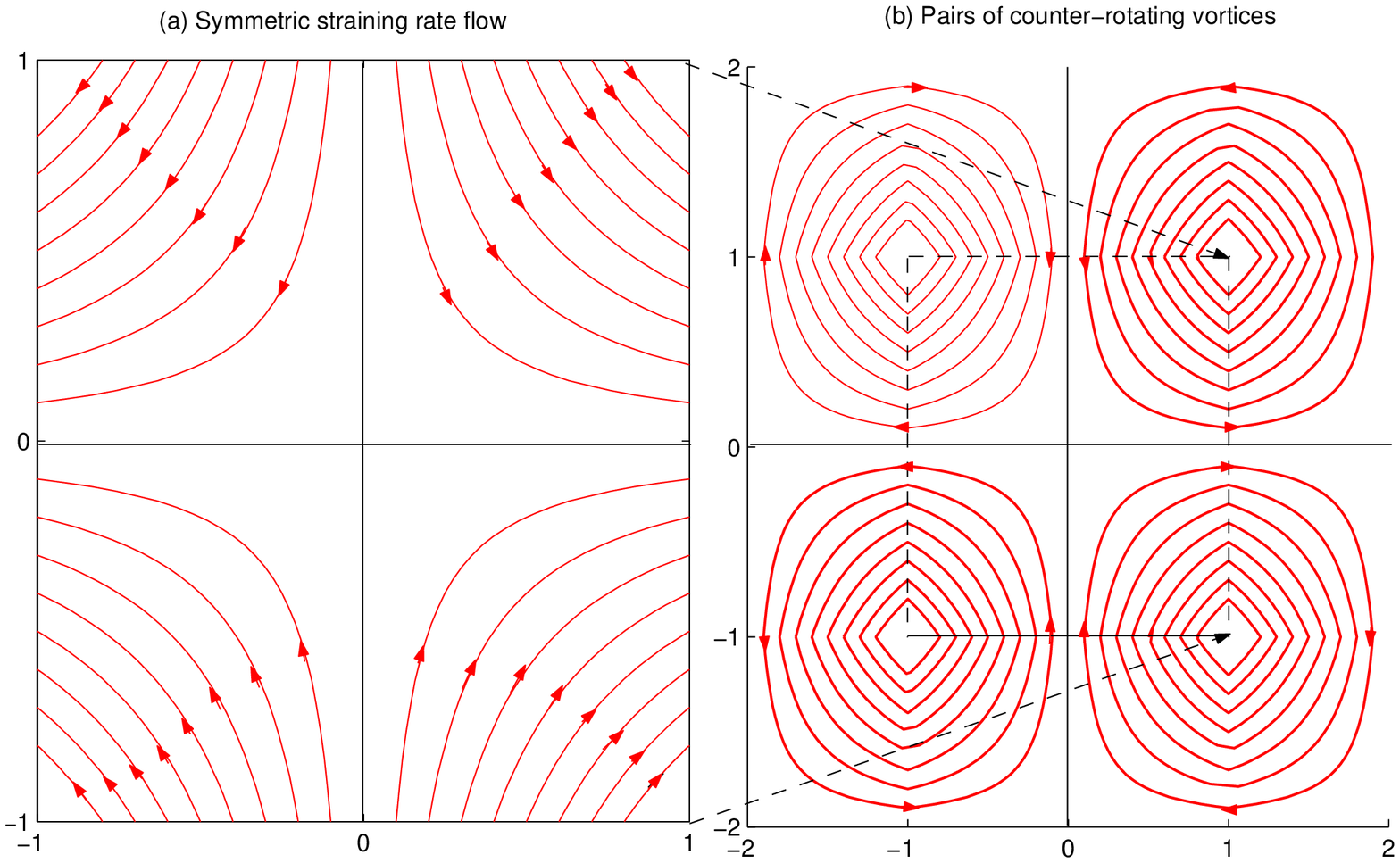}
 
\par\end{center}

\begin{center}
a) Generation from symmetric strain rate fl{}ow 
\par\end{center}%
\end{minipage}%
\begin{minipage}[c]{0.49\textwidth}%
\begin{center}
\includegraphics[scale=0.45]{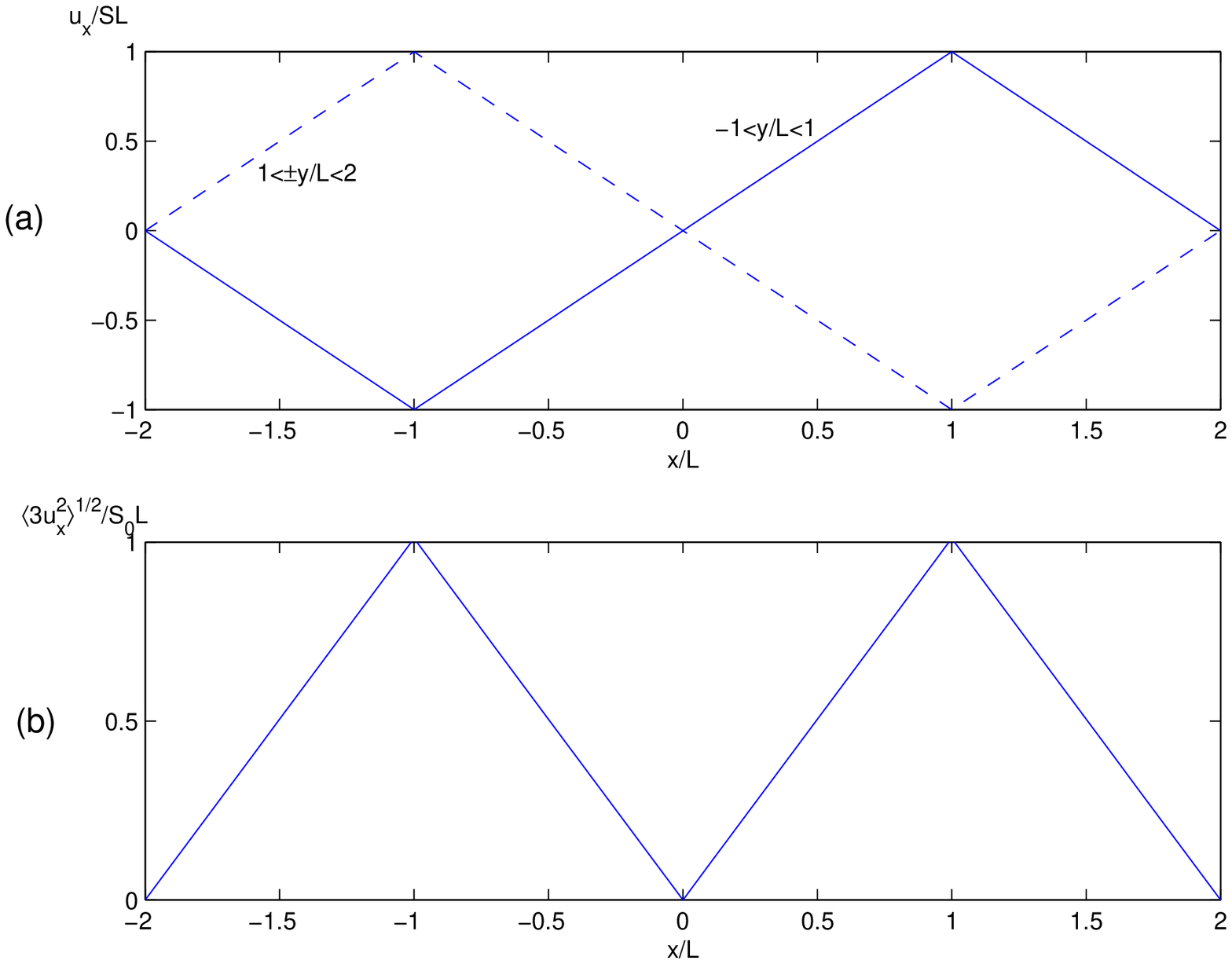}
 
\par\end{center}

\begin{center}
b) Carrier fl{}ow velocity and rms velocity in x-direction 
\par\end{center}%
\end{minipage}

\caption{\label{cap:Generation-of-carrier}Generation of carrier flow field
velocity and its $x-$variation within a flow lattice cell}
\end{figure}

This flow field and its properties we have described in detail elsewhere
\cite{key-54} but for completeness we shall include a brief description
here. We consider dispersion in a simple homogeneous turbulent flow
field composed of pairs of counter rotating vortices which are periodic
in both the \emph{x,y} directions with the same periodicity. Each
lattice cell (the basic periodic element) contains a pair of counter-rotating
vortices in both the \emph{x,y} orthogonal directions and is constructed
from a linear symmetric straining flow field in the manner shown in
Figure \ref{cap:Generation-of-carrier}

\end{multicols}

\begin{multicols}{2}

So starting from an initial symmetric straining flow pattern of width
$2L$ in both the $x,y$ directions (see Fig. \ref{cap:Generation-of-carrier}(a)
), this pattern is repeated \emph{front to back} in both the $x,y$
directions with a strain rate $S$ drawn from a uniform distribution
$\left[0,S_{0}\right]$. We note that each quadrant of this straining
rate pattern in Fig.\ref{cap:Generation-of-carrier}(a) is a quadrant
of one of the two pairs of counter-rotating vortices formed within
the lattice cell in Fig.\ref{cap:Generation-of-carrier}(b). As shown
in Fig.\ref{cap:Generation-of-carrier}(a), the flow velocity $u_{x}$
in the $x-$direction has a linear saw-tooth profile $U(x)$, with
a slope of constant magnitude $S$ but with a change in sign across
the $y$-centre line of a vortex %
\footnote{the line running in the y-direction passing through the centre of
the vortex%
} where the maximum and minimum values $\pm SL/2$ of $U(x)$ are located:
across the $x$-centre line, $u_{x}$ changes to $-U(x)$ as shown
in Fig.\ref{cap:Generation-of-carrier}(b), consistent with the change
in direction of the streamlines shown in Fig.\ref{cap:Generation-of-carrier}(a).
The flow velocity $u_{y}$ in the $y$-direction at $(x,y)$ is $-U(y)$
to preserve continuity of flow through out. This cellular flow pattern
of counter-rotating vortices so formed, persists for a fixed life-time
selected from an exponential distribution with a decay time of $S_{0}^{-1}$,
at the end of which time, a fresh flow field is generated with new
values of the life-time and $S$ and the origin of the pattern at
the same time shifted by random displacements in both the $x$ and
$y$-directions, drawn independently from a uniform distribution $\left[0,2L\right]$.
This makes the average flow homogeneous with zero mean in the $x-$
$y$-directions. The important feature of this randomized flow field
is that the equations of motion of an individual particle in both
the $x,y$-directions are linear and independent of one another (other
than through the maximum length of time a particle can experience
a particular value of the straining of the flow in either the $x$-
or $y$-directions before it changes sign). With respect to the centre
(stagnation point) of a symmetric straining flow pattern (see Fig.\ref{cap:Generation-of-carrier}(a)),
the flow velocity within that flow region is given by : 
\begin{equation}
u_{x}=+Sx\;;\; u_{y}=-Sy\;\;(-L\leq x\leq L,\:-L\leq y\leq L)
\end{equation}
 The flow field so generated turns out to be homogeneous and stationary
but not isotropic. It has the interesting property that the Lagrangian
fluid point rms velocity (along its trajectory) is different from
its value at a fixed point (Eulerian).

\subsection{Particle equations of motion }

\begin{figure}[htb]

\begin{centering}
\includegraphics[scale=0.7]{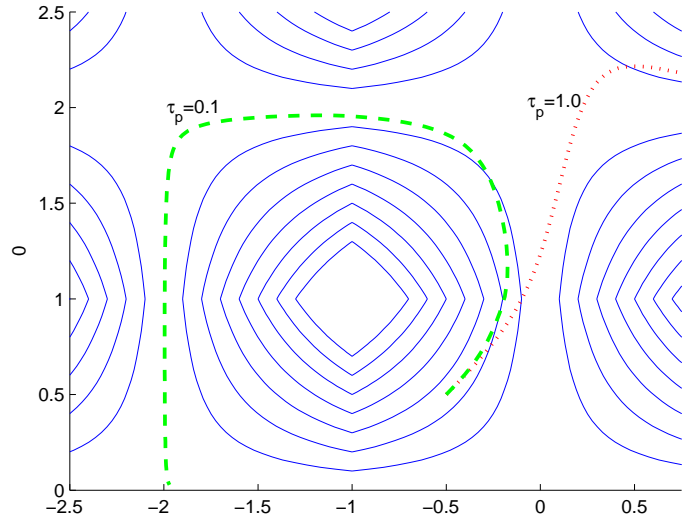} 
\par\end{centering}

\caption{\label{cap:Particle trajectories in vortices}Particle trajectories
in vortices: $\tau_{p}$=0.1 , heavily damped; $\tau_{p}=1,$lightly
damped}
\end{figure}

Based on Stokes drag, the particle equation of motion is 
\begin{eqnarray}
\ddot{x}_{i}+\tau_{p}^{-1}\dot{x}_{i}+(-1)^{i+1}\tau_{p}^{-1}Sx_{i} & = & 0\label{eq:SHM eqn}\\
-L\leq x_{i}\leq L\;(i=1,2)(x_{1}=x,x_{2}=y)\nonumber 
\end{eqnarray}
 where $x_{i}$ is measured from a stagnation point. For convenience
we normalize $x_{i}$ on $L$ and express the particle response time
$\tau_{p}$ and strain rate $S$ in units of $S_{0}^{-1}$ so $\tau_{p}$
is the particle Stokes number and in this case $S$ is drawn from
a uniform distribution $\wp\left[0,1\right]$. We note that when the
strain rate in the $x,y$ directions the $x,y$ equations are those
of a damped simple harmonic oscillator and as a consequence there
are two types of motion, namely heavily damped for $\tau_{p}<0.25$
and lightly damped for $\tau_{p}>0.25$. This has some bearing on
the motion of the particle within these vortices. For heavily damped
motion, a particle remains trapped within a vortex but approaches
the extremity (the stagnation region) in a manner which decreases
exponentially with time. On the other hand for lightly damped motion
the particle can escape the vortex or \emph{overshoot} into an adjacent
vortex, and in so doing overshoot into the adjacent vortex to that.
This has been referred to as the sling shot effect and it is this
type of motion that gives rise to RUM. Both these types of behaviour
are illustrated in Figure \ref{cap:Particle trajectories in vortices}
(a) and (b) for particle response times $\tau_{p}=0.1$ (heavily damped)
and $\tau_{p}=1$ (lightly damped).

\end{multicols}

\begin{multicols}{2}

\begin{figure}[htb]
\begin{minipage}[c]{0.49\textwidth}%
\begin{center}
\includegraphics[scale=0.5]{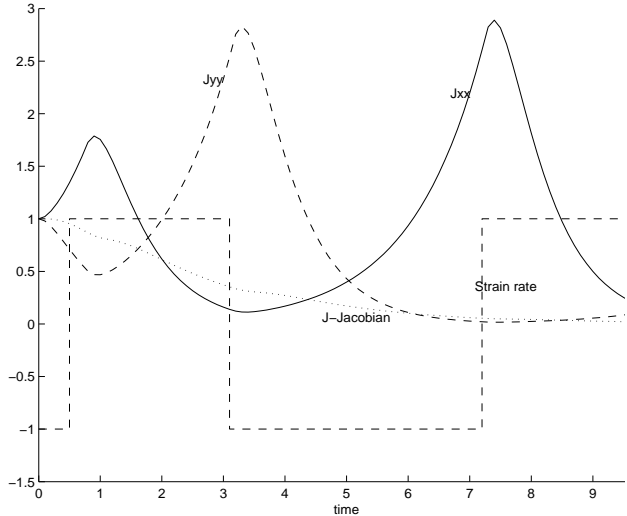}
 
\par\end{center}

\begin{center}
a) $St=0.1,\mbox{ heavilydamped}$ 
\par\end{center}%
\end{minipage}%
\begin{minipage}[c]{0.49\textwidth}%
\begin{center}
\includegraphics[scale=0.4]{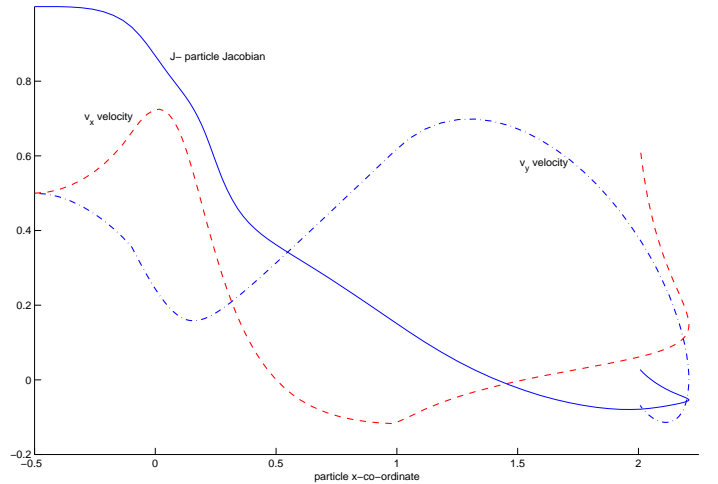}
 
\par\end{center}

\begin{center}
b) $St=1$., lightly damped 
\par\end{center}%
\end{minipage}

\caption{\label{cap:Compenents-of-J}Compenents of the deformation tensor $\mathbf{J}$
and compression $\left|J\right|$along particle trajectories : }
\end{figure}

Figure\ref{cap:Compenents-of-J} also show the corresponding values
of components $J_{11}$ and $J_{22}$ of the the unit deformation
tensor $\mathbf{J}$ given by $\partial\mathbf{x}(\mathbf{x_{0},}t)/\partial\mathbf{x}_{0}$
where \textbf{$\mathbf{x_{0}}$} is the position of the particle at
some initial time say $t=0$. In this case we have set $J_{ij}(0)=\delta_{ij}$.
The equations of motion for $J_{ij}$ are obtained from the particle
equations of motion by partial differentiation wrt \textbf{$x_{0,i}$}
giving 
\begin{equation}
\ddot{J_{ij}}+\tau_{p}^{-1}\dot{J}_{ij}+(-1)^{i+1}\tau_{p}^{-1}SJ_{ij}=0\label{eq:}
\end{equation}
 for which we choose the initial conditions $\dot{J}_{ij}=\partial u_{i}/\partial x_{j}=(-1)^{i+1}\tau_{p}^{-1}S\delta_{ij}$.
This means that together with the initial conditions on $J_{ij}$
that $J_{ij}(t)=0$ for $i\neq j$ and that 
\begin{equation}
J(t)=\left|\mathbf{J}\right|=J_{11}J_{22}\label{eq:}
\end{equation}

\end{multicols}

\begin{multicols}{2}

Note that in this linear system the equation of motion for $J(t)$
is the same as that for $x_{i}$ and that the components of $\mathbf{J}$
are dependent on the position of the particle only through the time
at which the strain rate experienced by the particle changes sign.
Fig \ref{cap:Compenents-of-J}(a) and (b) shows the corresponding
values of $J(t)$ along the heavily damped and lightly damped (b)
trajectories. In both cases the value of $J$ approaches zero as $t\rightarrow\infty$.
However in the lightly damped case $J(t)$ passes through zero at
intermediate times as the particle oscillates backwards and forwards
across a stagnation line. In so doing the value of $J$ oscillates
from $+ve$ to $-ve$, with the corresponding elemental volume rotating
through $180^{0}$ as it passes through zero volume. Each time $J(t)$
passes through zero, the corresponding particle concentration becomes
infinite instantaneously. This raises the possibility that such events
may occur in real turbulent flows and that the process of particle
dispersion could be a highly intermittent process associated with
large deviations in the particle concentrations. It also indicates
the conditions under which particle trajectories cross. That is given
an initially uniform distribution of particle with velocities uniquely
defines at each point by the local fluid velocity., as time elapses
the particle velocity field is not uniquely defines at each point.
This ultimately is the origin of RUM.

\section{Calculation of RUM}

We calculated RUM by measuring the contribution it makes to the particle
kinetic energy (per unit mass). We did this by calculating the particle
pair velocity correlation as a function of particle separation. In
particular we calculated 
\begin{equation}
R_{xx}(\Delta x)=\langle\textrm{v}_{x}^{(1)}(x,0)\textrm{v}_{x}^{(2)}(x+\Delta x,0)\rangle/\langle\textrm{v}_{x}^{2}\rangle\label{eq:pair velocity correlation}
\end{equation}
 where $(1)$and $(2)$refer to two particles labelled (1) and $(2)$separated
by a distance $\left[\Delta x,0\right]$and $\langle\rangle$is an
average over all particle pairs for which the separation is $\Delta x$.
We began with a fully mixed flow of particles with the local fluid
velocity. Periodicity of the lattice cell dimensions means that initially
for any integers $n$, and $m$,the initial particle velocity field
satisfies the periodic boundary condition, $\mathbf{v}(x,y)=\mathbf{v}(x+4n,y+4m)$.
Given this initial velocity field means that for any given particle
trajectory $\mathbf{X}_{1}(\mathbf{x}_{1},t)$ there will be an identical
particle trajectory $\mathbf{X}_{2}(x_{1}+4n,y+4m_{1})$, for which
\begin{equation}
\mathbf{X}_{2}(x_{1}+4n,y+4m_{1},t)=\mathbf{X}_{1}(\mathbf{x_{1}},t)+\mathbf{\Gamma};\:\mathbf{\Gamma}=(4n,4m)\label{eq:}
\end{equation}
 That is two particle starting out from identical positions within
any two lattice cells will experience the same relative displacement
with respect to their initial positions and necessarily experience
the same flow velocity at any given time even though the whole lattice
is displaced randomly at each life time. This means that if we consider
particles entering or leaving a fixed cell of lattice dimensions,
any particle leaving the cell will be replaced by an identical particle
at the opposite face whose velocity is identical to that of the particle
leaving. Thus at any instant of time, the number of particles within
the cell remains constant; only the concentration within the cell
will change. Periodicity of the particle motion and homogeneity means
further more that for the particle pair velocity correlation 
\begin{equation}
\begin{array}{c}
R_{xx}(\Delta x)=R_{xx}(-\Delta x)=R_{xx}(4-\Delta x)=R_{yy}(\Delta y)\\
=R_{yy}(-\Delta y)=R(4-\Delta y)
\end{array}\label{eq:}
\end{equation}
 Because the particle concentration is initially uniform, the particle
pair correlation is identical to the Eulerian spatial velocity correlation
which we can calculate directly, namely

\end{multicols}

\begin{equation}
R_{xx}(\Delta x)=\int_{lattice}u_{x}(x,y)u_{x}(x+\Delta x,y)dxdy/\langle u_{x}^{2}\rangle=\frac{1}{16}\int_{-1}^{3}\int_{-1}^{3}u_{x}(x,y)u_{x}(x+\Delta x,y)dxdy\label{eq:}
\end{equation}

\begin{figure}[htb]

\begin{centering}
\includegraphics[angle=270,scale=0.5]{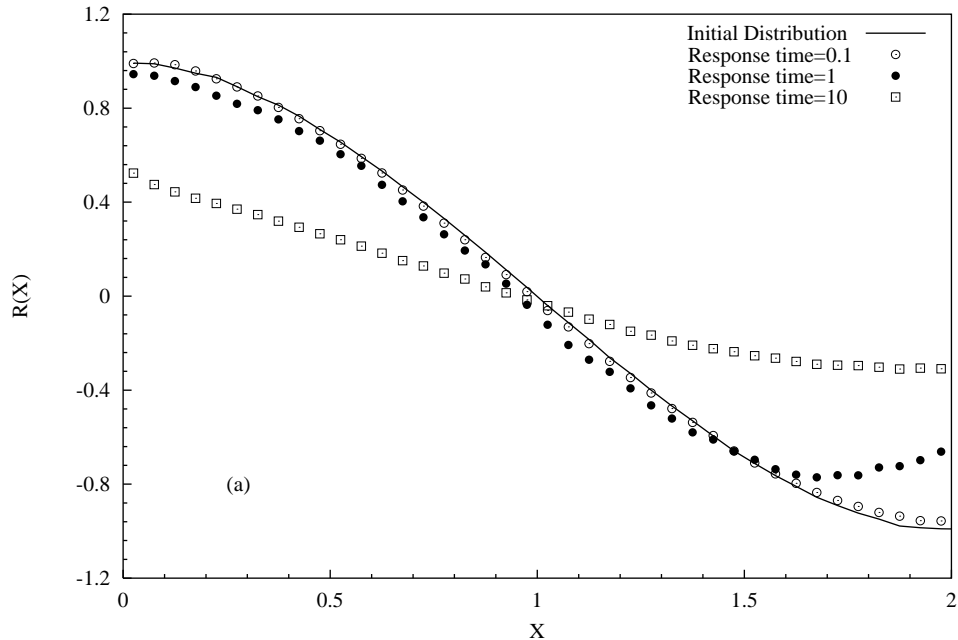} 
\par\end{centering}

\caption{\label{correlation function t=00003D00003D00003D3D20}correlation
coefficient $R_{xx}(X)$ at time $t=40$, where $X$is the $x$-separation
of particle pairs; note that , $R_{xx}(X)$is symmetric $X=2$}
\end{figure}

After some labour, one arrives at the result 
\begin{equation}
\begin{array}{cc}
0\leq\Delta x\leq2 & \; R_{xx}(\Delta x)=\left[1-\frac{3}{2}\Delta x^{2}+\frac{1}{2}\Delta x^{3}\right]\\
2\leq\Delta x\leq4 & R_{xx}(\Delta x)=\frac{9}{12}\left[12-16\Delta x+6\Delta x^{2}-\frac{2}{3}\Delta x^{3}\right]=\left[9-12\Delta x+\frac{9}{2}\Delta x^{2}-\frac{1}{2}\Delta x^{3}\right]
\end{array}\label{eq:correlation for initail separation}
\end{equation}
 These forms provide an independent check on our method of calculating
the particle pair velocity correlation based on counting the number
of particle pairs with a given separation . The method within the
statistical error correctly reproduces the correlation coefficient
for the particle initial uniform distribution (See Fig. \ref{correlation function t=00003D00003D00003D3D20})

\begin{figure}[htb]
\begin{minipage}[c]{0.49\textwidth}%
\begin{center}
\includegraphics[angle=270,scale=0.3]{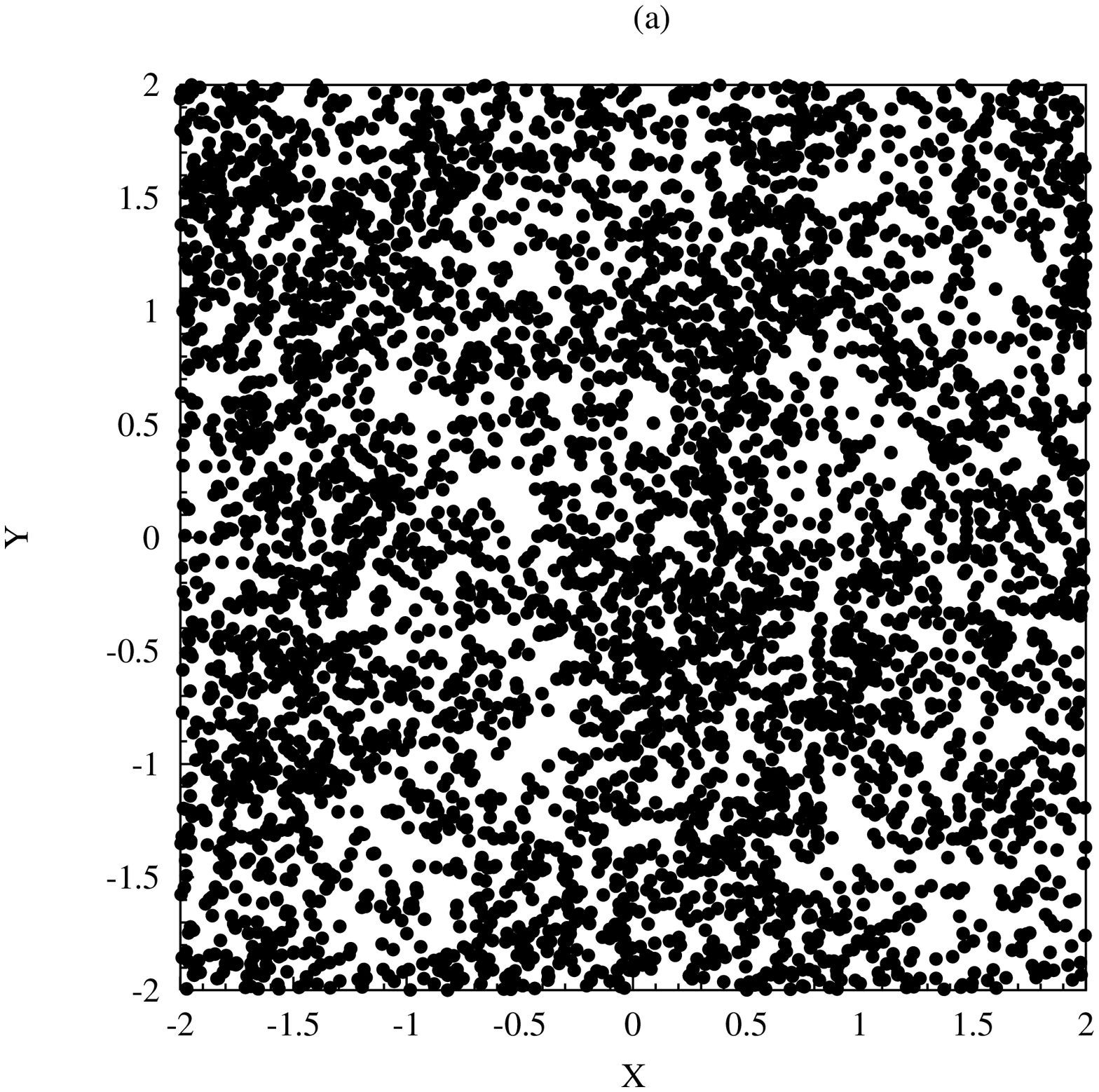}
 
\par\end{center}

\begin{center}
a) St=0.1, time =10 
\par\end{center}%
\end{minipage}%
\begin{minipage}[c]{0.49\textwidth}%
\begin{center}
\includegraphics[angle=270,scale=0.25]{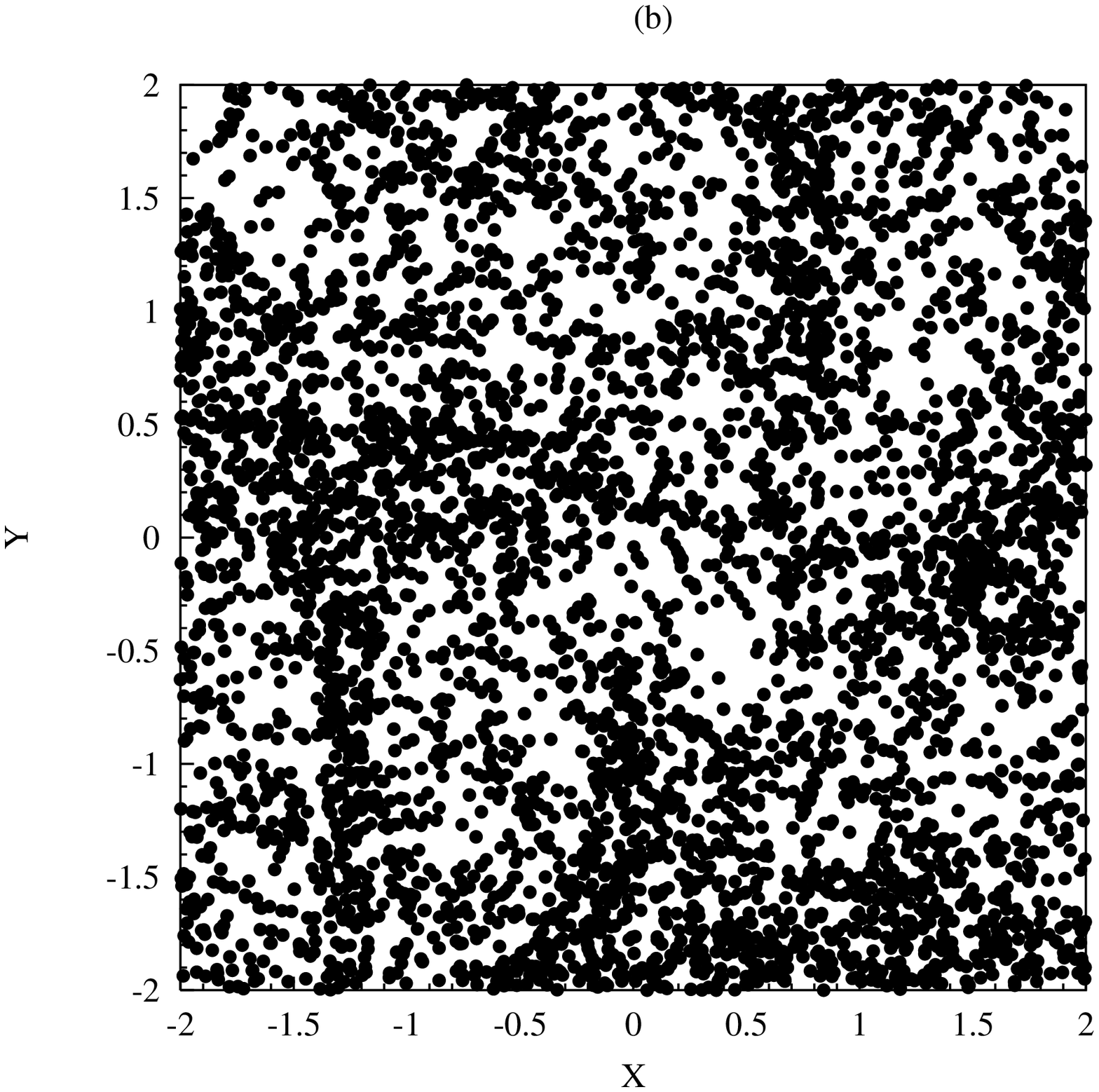}
 
\par\end{center}

\begin{center}
b) St=0.1, time=40 
\par\end{center}%
\end{minipage}\\
\begin{minipage}[c]{0.49\textwidth}%
\begin{center}
\includegraphics[angle=270,scale=0.3]{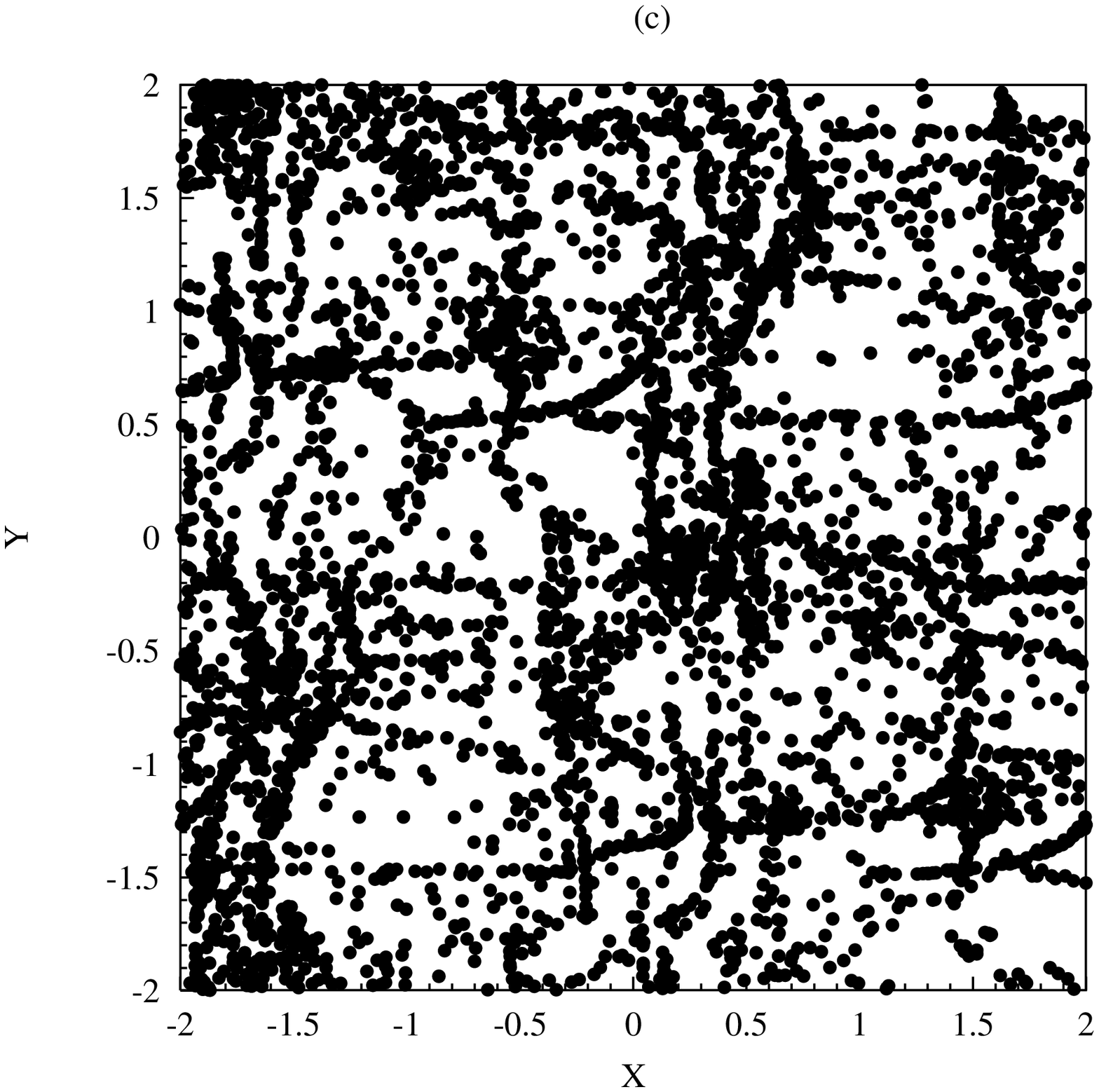}
 
\par\end{center}

\begin{center}
c) St=1, time=10 
\par\end{center}%
\end{minipage}%
\begin{minipage}[c]{0.49\textwidth}%
\begin{center}
\includegraphics[angle=270,scale=0.3]{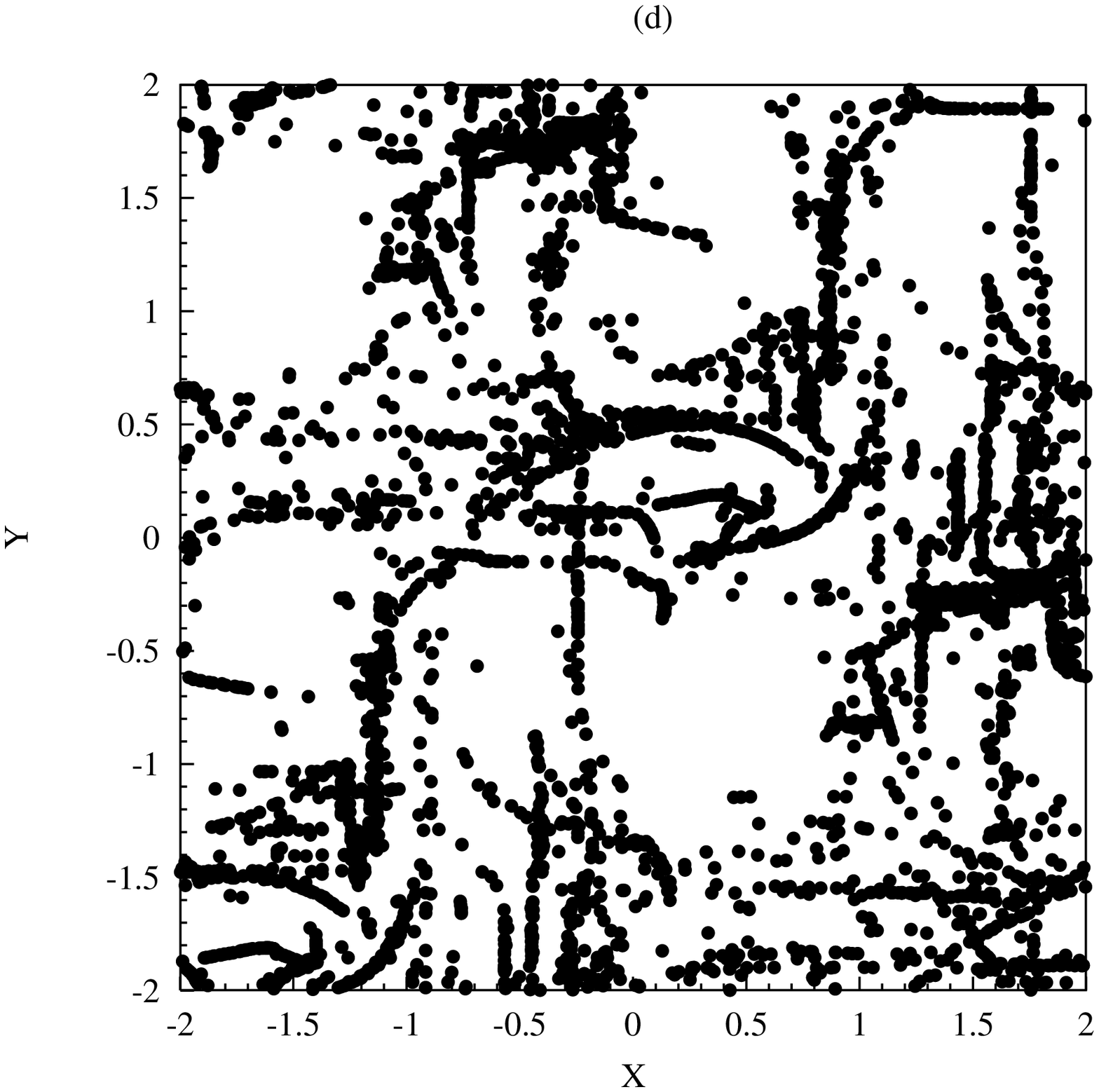}
 
\par\end{center}

\begin{center}
d) St=1, time=40 
\par\end{center}%
\end{minipage}\\
\begin{minipage}[c]{0.49\textwidth}%
\begin{center}
\includegraphics[angle=270,scale=0.3]{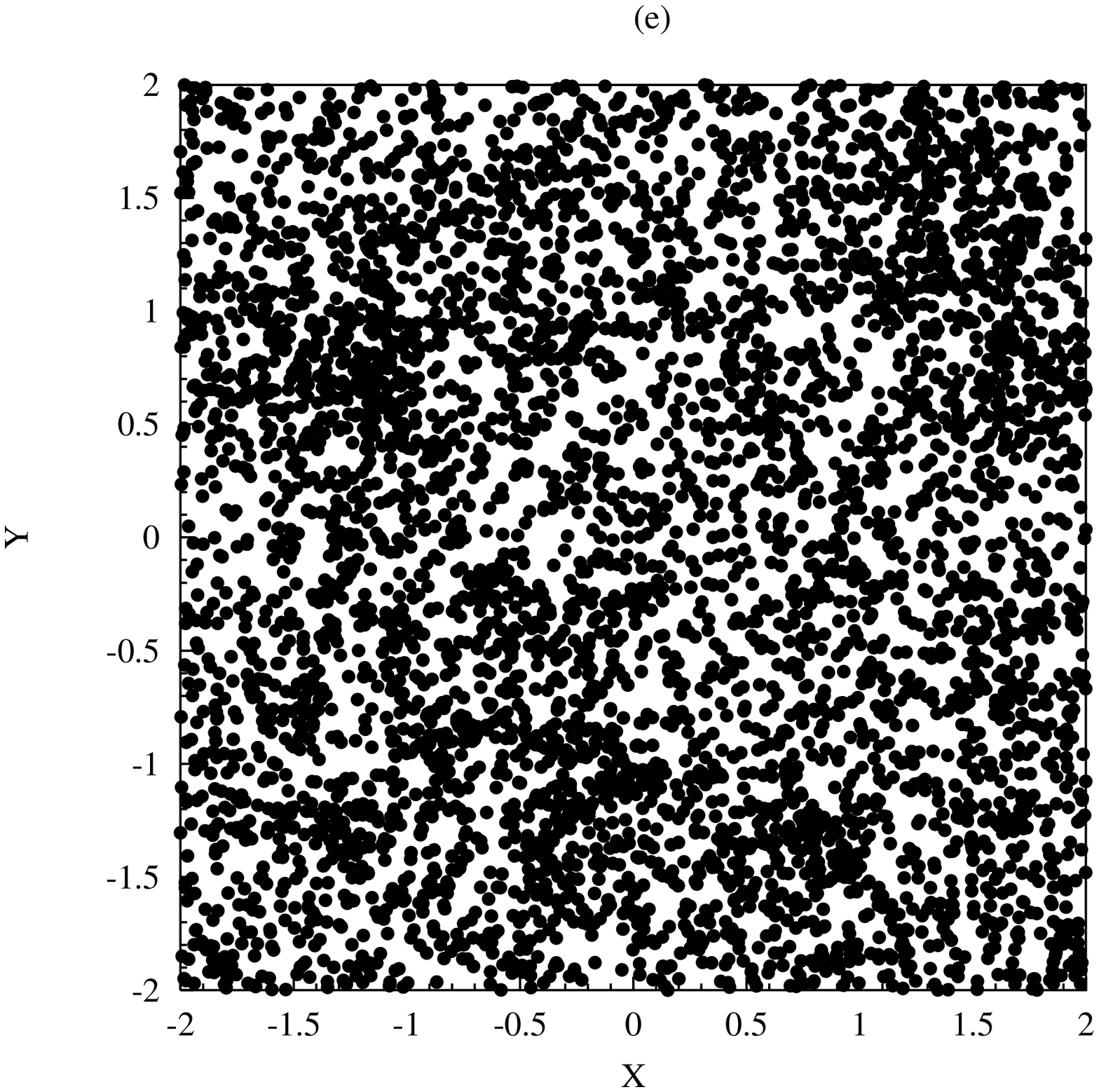}
 
\par\end{center}

\begin{center}
e) St=10, time=10 
\par\end{center}%
\end{minipage}%
\begin{minipage}[c]{0.49\textwidth}%
\begin{center}
\includegraphics[angle=270,scale=0.3]{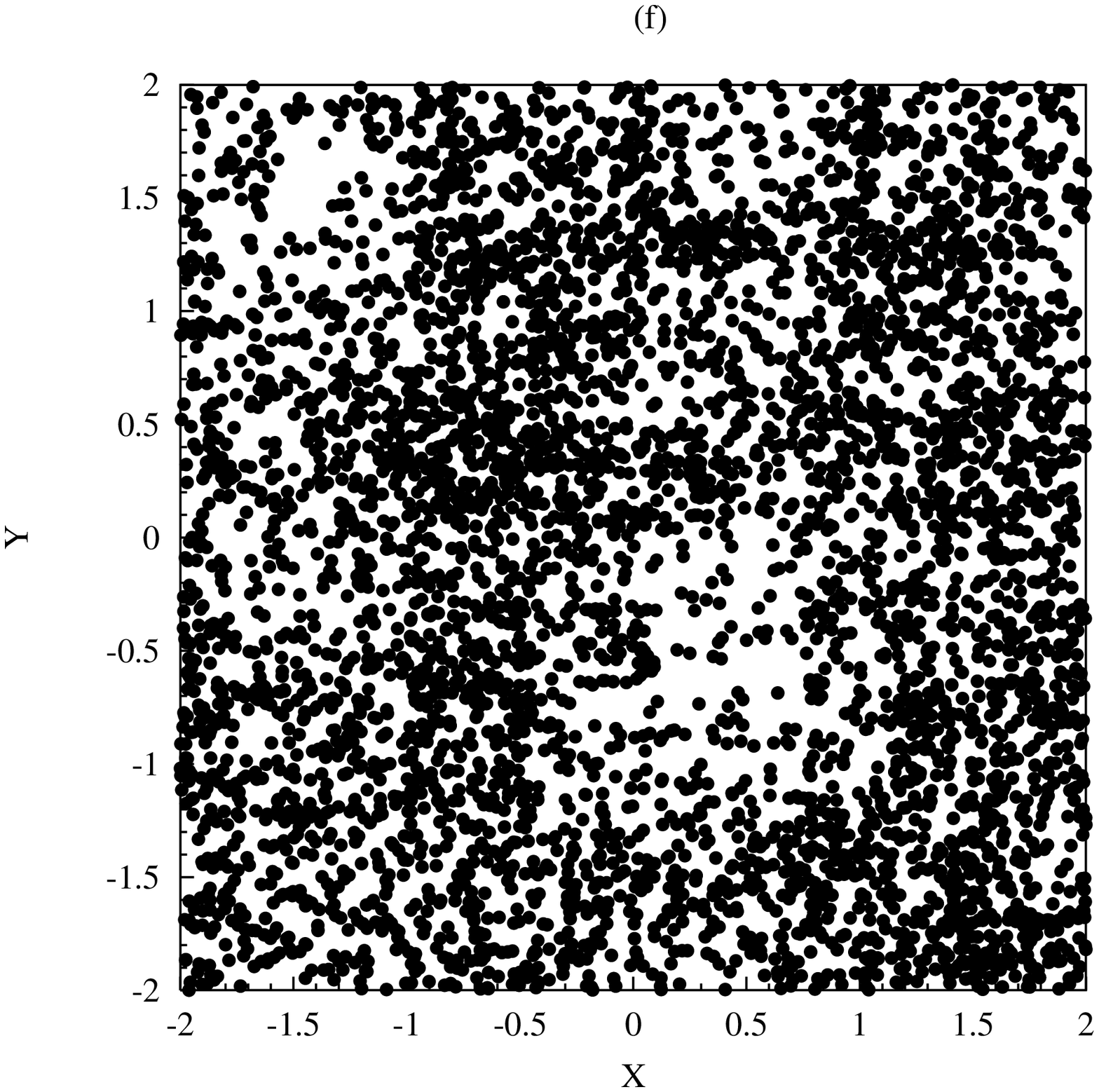}
 
\par\end{center}

\begin{center}
f) St=10, time=40 
\par\end{center}%
\end{minipage}

\caption{\label{Particle distribitution}Particle distributions at time $t=10$
and 40, for $St=0.1,1.,10$ }
\end{figure}

\begin{multicols}{2}

Figure \ref{Particle distribitution} shows the positions of $500$
particles initially distributed uniformally within a cell whose boundaries
are coincident with a lattice cell initially but which remains fixed
throughout the simulation, and for which periodic boundary conditions
are applied at the edges of the cell to particles entering and leaving
it. The resulting particle distributions in Figure\ref{Particle distribitution}
are for a given random sequence of the straining rate$S$and \emph{lattice}
life time. The particular distributions shown are at times $t=10$
and $t=40$ for particle response times (Stokes nos.) of $\tau_{p}=0.1,1.0,10.$
We note first that the case of the intermediate response time $\tau_{p}=1$,
exhibits the greatest degree of segregation, being markedly greater
than the other two cases. Secondly the segregation increases with
time and carries on indefinitely. Thirdly there is no identifiable
self preserving pattern of segregation, nor does the segregation align
with the local stagnation lines of the flow at any instant of time.
In fact another random sequence of strain rates and lifetimes would
lead to an entirely different pattern. The increasing segregation
would lead to a lower fractal dimension and the increase of the particle
pair concentration at zero separation values for the pair separation
distribution function as shown in Figure \ref{Particle distribitution}
described below.

In calculating the correlation coefficient $R_{xx}(\Delta x)$for
particle pair separation in the $x$-direction, the positions of $N$
particles within the domain $\left[0\leq\Delta x\leq4\,,\,0\leq y\leq4\right]$are
evaluated as a function of time by first solving the set of equations
of motion (\ref{eq:SHM eqn}). The domain $\left[0\leq\Delta x\leq4\,,\,0\leq y\leq4\right]$is
divided up into $10^{2}\times10^{2}$ bins of $0.04$ dimension in
both the $\Delta x,y$ directions, the $(m,n)$bin being the bin whose
centre is at $\Delta x=$$0.04(m-\frac{1}{2}),y=0.04(n-\frac{1}{2})$
where $m,n=1,..100$. For each particle pair whose $y$- co-ordinates
satisfy $0.04(n-1)<y<n0.04$ we evaluate the separation $\Delta x$.
If $\Delta x>0$ we determine the value of $m$ for which $0.04(m-1)<\Delta x<m0.04$
and store the value $\textrm{v}_{x}^{(1)}(x,0)\textrm{v}_{x}^{(2)}(x+\Delta x,0)$
in the $(m,n)$bin. If $\Delta x<0$, we store these values in the
$\left(m,n\right)$bin for which $0.04m<4-\Delta x<(m+1)0.04$. In
this way, for all particle pairs whose $x-$position is $-2<x<2$
we cover the range $0\leq\Delta x\leq4$. After sorting all particle
pairs into suitable bins, we then repeat this procedure for $M$ realisations
of the carrier flow and then sum the values of $\textrm{v}_{x}^{(1)}(x,0)\textrm{v}_{x}^{(2)}(x+\Delta x,0)$for
all particle pairs within each bin and divide this by the number of
particle pairs to give the average value $\left\langle \textrm{v}_{x}^{(1)}(x,0)\textrm{v}_{x}^{(2)}(x+\Delta x,0)\right\rangle $.
In addition we calculate $\langle\textrm{v}_{x}^{2}\rangle$by summing
$\textrm{v}_{x}^{2}$over all $N$ particles in the box and over $M$
realisations of the flow field and then dividing by $MN,$ finally
to obtain $R_{xx}(\Delta x)$ using Eq.(\ref{eq:pair velocity correlation}).
For accuracy versus computation time we chose $N=10^{3}$ particles
and $M=10^{3}$ realisations of the flow. At the same time as summing
over the values $\textrm{v}_{x}^{(1)}(x,0)\textrm{v}_{x}^{(2)}(x+\Delta x,0)$
in all bins with the same value of $n$, we also summed the number
of pairs in each bin and divided by $0.04$, to evaluate the particle
pair distribution $g(\textrm{\ensuremath{\Delta x}})$which we normalised
to unity by dividing the number of pairs within the domain $0\leq\Delta x\leq4$.
i.e. 
\[
\int_{0}^{4}g(\Delta x)d\Delta x=1
\]

\begin{figure}[htb]
\begin{centering}
\includegraphics[angle=270,scale=0.5]{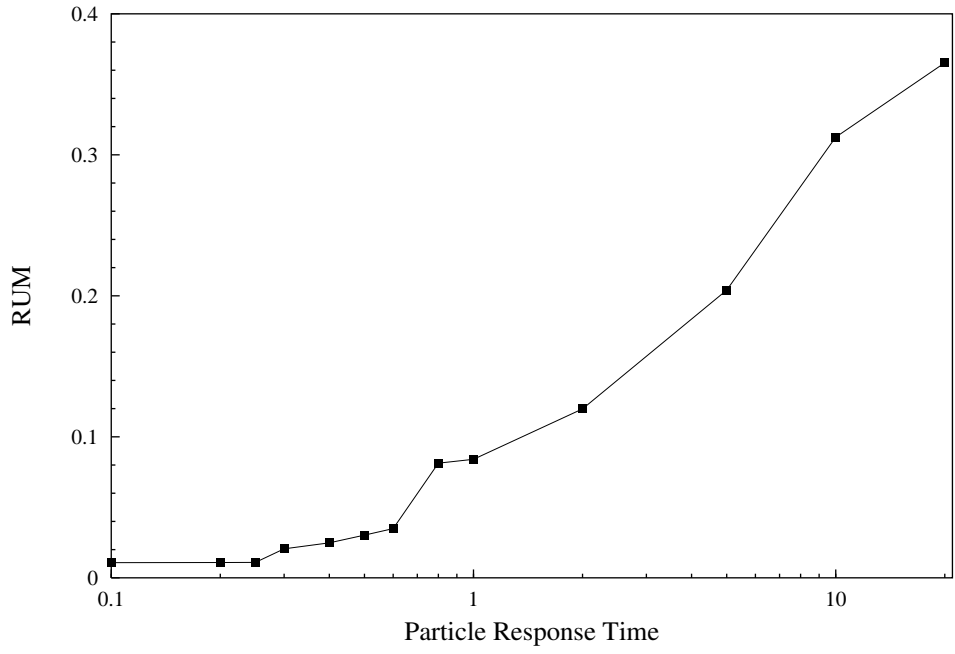} 
\par\end{centering}

\caption{\label{RUM contribution to Particle KE}RUM contribution to particle
kinetic energy}
\end{figure}

For each of values of $\tau_{p}=0.1,1,10$, Figure \ref{correlation function t=00003D00003D00003D3D20}
shows the particle pair velocity correlation coefficient $R_{xx}(\Delta x)$
at time $t=20)$ as a function of the pair separation. These three
curves are to be compared with the curve for the initial distribution
given by Eq.(\ref{eq:correlation for initail separation}). It is
apparent that the case for $\tau_{p}=0.1$ is almost identical to
the initial form of the pair separation correlation. For the other
cases, the curves are noticeably different, with the most interesting
property of all, that the extrapolated values for $\Delta x=0$are
not unity as is the case of the initial pair correlation function
and that of the case of $\tau=0.1$. This is most marked for $\tau_{p}=10$.
The difference between the extrapolated value and unity represents
the contribution of RUM to the particle kinetic energy. More revealing
are the calculated values of RUM plotted against $\tau_{p}$ shown
in Fig.\ref{RUM contribution to Particle KE} for $t=20$. We note
that there is no real measurable contribution from RUM until $\tau_{p}\geq0.25$.
For $\tau_{p}\leq0.25$ all realisable states correspond to a sequence
of heavily damped system with $\tau_{p}S$$\leq0.25$: above $\tau_{p}=0.25$
the system behaves as a mixture of heavily damped and lightly damped
harmonic oscillators. The threshold of $\tau_{p}=0.25$ for the occurrence
of RUM, shows conclusively that RUM is associated with the behaviour
of a lightly damped system in the manner described earlier.

\begin{figure}[htb]

\begin{centering}
\includegraphics[scale=0.3,angle=270]{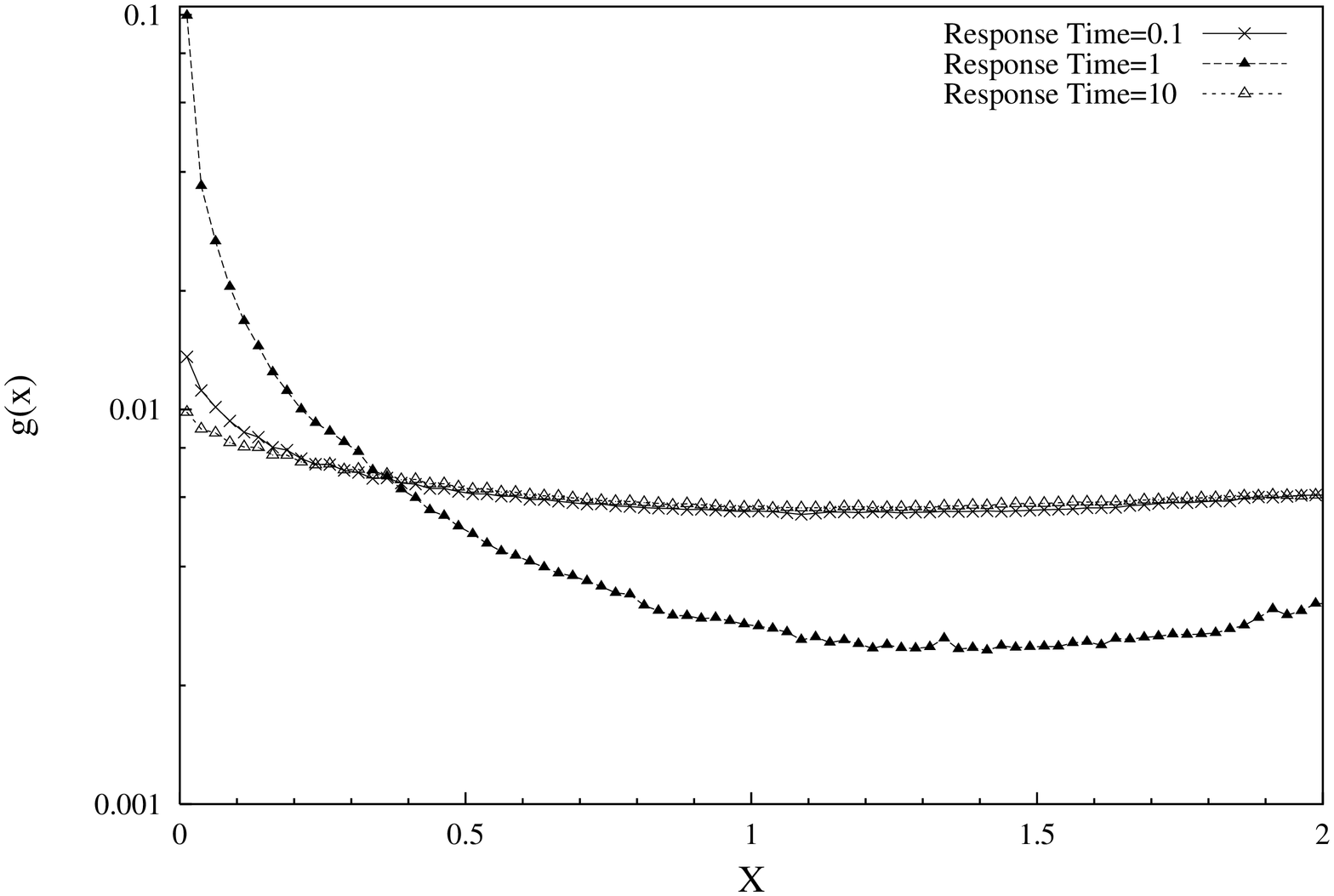}\includegraphics[scale=0.3,angle=270]{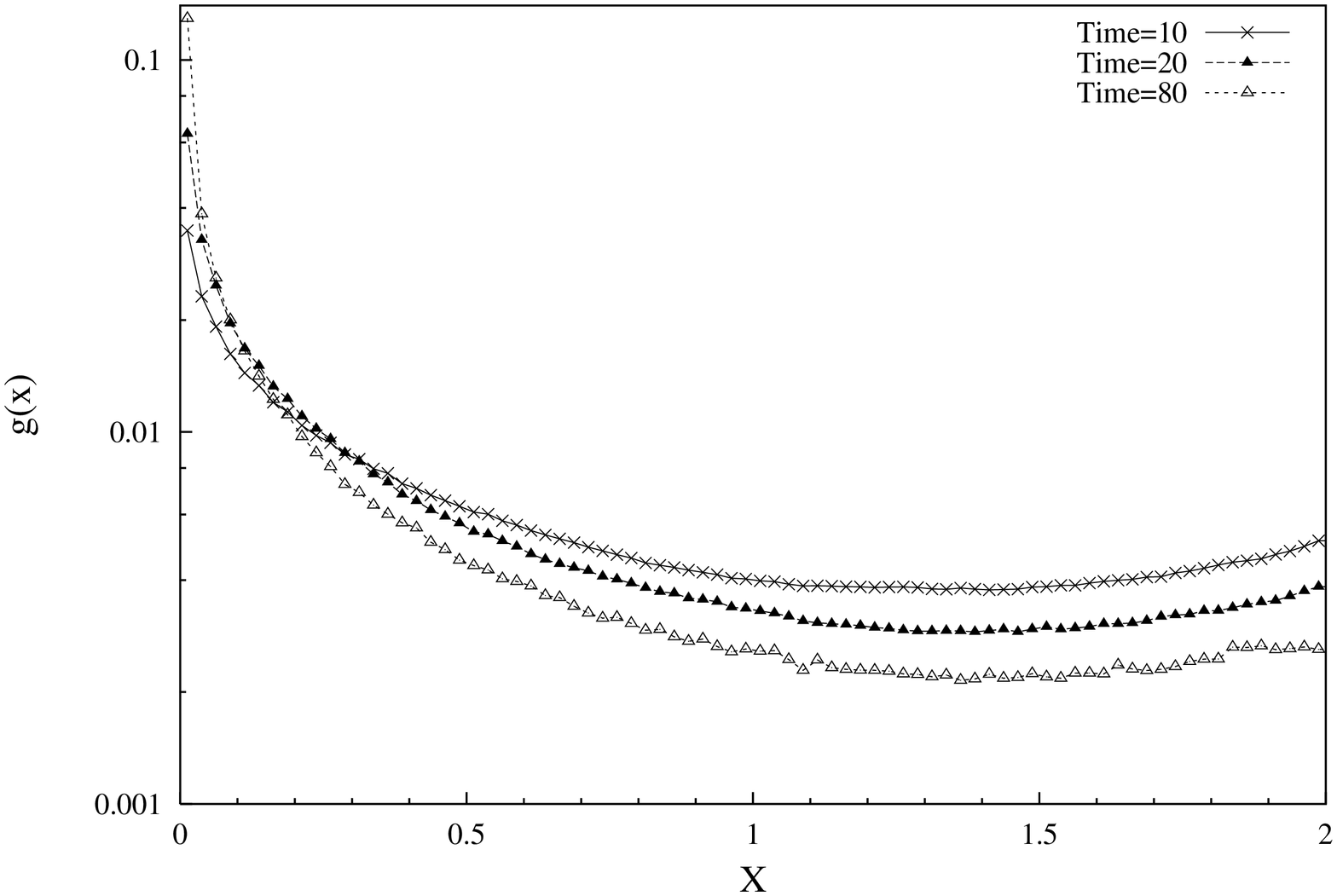} 
\par\end{centering}

\caption{\label{cap:Partice-pair-distribution}Particle pair distribution $g(X)$,
$X$= particle pair separation}
\end{figure}

Finally Figure \ref{cap:Partice-pair-distribution} show the calculated
values of the particle pair distribution function $g(x)$. Figure
\ref{cap:Partice-pair-distribution} (a) shows that for the 3 values
of $\tau_{p}=0.1,1.0,10$ the maximum peak concentration $\Delta x=0$occurs
at the intermediate values of $\tau_{p}=1.$. Fig\ref{cap:Partice-pair-distribution}(b)
shows peak distribution increasing with time for $\tau_{p}=1,$a result
that is true for all values of $\tau_{p}$. We recall that these features
are consistent with the segregation of particles shown in Figure \ref{Particle distribitution}

\section{Summary and Concluding remarks}

We have shown and measured a contribution to the particle kinetic
energy from a component of random uncorrelated motion (RUM) of particles
suspended in a simple random but smoothly varying flow field composed
of counter rotating vortices. This feature has been observed in more
complex flow. Depending upon the value of the particle response time
$\tau_{p}$(Stokes number) the motion of the particles corresponds
to that of a heavily or lightly damped harmonic oscillator. Within
statistical error, RUM was only detected when the particle motion
at some stage corresponded to a lightly damped oscillator in particular
when $\tau_{p}\geq0.25$. In this case the particles overshoot the
stagnation lines in the flow and can pass form one vortex to another
in the flow. This behaviour is consistent with the description given
by Simonin et al.\cite{key-1} who first observed RUM in particle
motion in DNS isotropic homogeneous turbulence. There remains the
possibility of linking RUM with the occurrence of singularities in
the particle concentration flow field and to the intermittency in
the particle flow filed. Simonin et al. \cite{key-1} have developed
a statistical approach to describing particle motion in these circumstances.
They have considered RUM as a device for stabilising the segregation
(the RUM acting as a pressure, the gradient of which acts in opposition
to the force driving the segregation. Here we note that the segregation
does not stabilise even though RUM is definitely present.

\end{multicols}

\begin{multicols}{2} 

\end{multicols}

\end{document}